\documentclass{jpc} 
\usepackage[utf8]{inputenc}
\usepackage{amsmath, amsfonts, amssymb, amsthm}
\usepackage{mathtools}
\usepackage{enumitem}
\usepackage{url}
\usepackage{hyperref}
\usepackage{graphicx}
\usepackage{comment}
\usepackage{caption} 
\usepackage{subcaption}
\usepackage{float}
\usepackage{xspace}
\usepackage{suffix}
\usepackage{makecell}
\usepackage{balance}  
\usepackage[numbers]{natbib} 
\usepackage{booktabs}
\usepackage{multirow}
\usepackage[linesnumbered,lined,ruled,vlined]{algorithm2e}
\makeatletter
\newcommand{\nosemic}{\renewcommand{\@endalgocfline}{\relax}}
\newcommand{\dosemic}{\renewcommand{\@endalgocfline}{\algocf@endline}}
\let\oldnl\nl
\newcommand{\nonl}{\renewcommand{\nl}{\let\nl\oldnl}}
\makeatother
\def\ddefloop#1{\ifx\ddefloop#1\else\ddef{#1}\expandafter\ddefloop\fi}
\def\ddef#1{\expandafter\def\csname bb#1\endcsname{\ensuremath{\mathbb{#1}}}}
\ddefloop ABCDEFGHIJKLMNOPQRSTUVWXYZ\ddefloop
\def\ddef#1{\expandafter\def\csname c#1\endcsname{\ensuremath{\mathcal{#1}}}}
\ddefloop ABCDEFGHIJKLMNOPQRSTUVWXYZ\ddefloop
\def\ddef#1{\expandafter\def\csname v#1\endcsname{\ensuremath{\boldsymbol{#1}}}}
\ddefloop ABCDEFGHIJKLMNOPQRSTUVWXYZabcdefghijklmnopqrstuvwxyz\ddefloop
\def\ddef#1{\expandafter\def\csname v#1\endcsname{\ensuremath{\boldsymbol{\csname #1\endcsname}}}}
\ddefloop {alpha}{beta}{gamma}{delta}{epsilon}{varepsilon}{zeta}{eta}{theta}{vartheta}{iota}{kappa}{lambda}{mu}{nu}{xi}{pi}{varpi}{rho}{varrho}{sigma}{varsigma}{tau}{upsilon}{phi}{varphi}{chi}{psi}{omega}{Gamma}{Delta}{Theta}{Lambda}{Xi}{Pi}{Sigma}{varSigma}{Upsilon}{Phi}{Psi}{Omega}{ell}\ddefloop
\makeatletter
\DeclareRobustCommand\widecheck[1]{{\mathpalette\@widecheck{#1}}}
\def\@widecheck#1#2{%
    \setbox\z@\hbox{\m@th$#1#2$}%
    \setbox\tw@\hbox{\m@th$#1%
       \widehat{%
          \vrule\@width\z@\@height\ht\z@
          \vrule\@height\z@\@width\wd\z@}$}%
    \dp\tw@-\ht\z@
    \@tempdima\ht\z@ \advance\@tempdima2\ht\tw@ \divide\@tempdima\thr@@
    \setbox\tw@\hbox{%
       \raise\@tempdima\hbox{\scalebox{1}[-1]{\lower\@tempdima\box
\tw@}}}%
    {\ooalign{\box\tw@ \cr \box\z@}}}
\makeatother
\DeclarePairedDelimiter\ang{\langle}{\rangle}  
\DeclarePairedDelimiter\abs{\lvert}{\rvert}    
\DeclarePairedDelimiter\norm{\lVert}{\rVert}   
\DeclarePairedDelimiter\set{\{}{\}}            
\DeclarePairedDelimiter\floor{\lfloor}{\rfloor}
\newcommand{\pr}[1]{\ensuremath{P\left(#1\right)}}  
\newcommand{\one}[1]{\ensuremath{\mathbf{1}_{#1}}}  
\DeclareMathOperator{\bern}{Bernoulli}          
\DeclareMathOperator{\lap}{Lap}                 
\DeclareMathOperator{\expo}{Exp}               

\DeclareMathOperator{\unif}{\mathcal{U}}          
\DeclareMathOperator{\geom}{Geo}               

\newcommand{\down}[2][]{\floor{{#2}}_{{#1}}}       
\WithSuffix\newcommand\down*[2][]{\floor*{#2}_{#1}} 
\newcommand{\round}[2][]{\ang{#2}_{#1}}        
\WithSuffix\newcommand\round*[2][]{\ang*{#2}_{#1}}  
\newcommand{\tgtres}{\ensuremath{\gamma_{*}}}
\newcommand{\res}[1][]{{\ensuremath{\gamma}_{#1}}}
\newcommand{\rat}{\theta}
\DeclareMathOperator{\remainder}{rem}
\newcommand{\mech}{\mathcal{M}}
\newcommand{\noisymax}{Noisy Max\xspace}

\newcommand{\topk}{Noisy Top-k\xspace}
\newcommand{\gaptopk}{Noisy Top-k with Gap\xspace}

\usepackage{xargs}

\def\notationcolor{black} 

\newcommand{\notation}[2]{\newcommand{#1}{{\textcolor{\notationcolor}{\ensuremath{#2}}}}}

\notation{\exprv}{X} 
\notation{\georv}{Y} 
\notation{\tgeorv}{Y'} 
\notation{\texprv}{X'} 
\notation{\sgeorv}{Z}

\notation{\data}{D}

\keywords{differential privacy, floating-point vulnerability, report noisy top-$k$}

\usepackage{natbib}
\usepackage[ruled]{algorithm2e}
\theoremstyle{plain} 
\theoremstyle{plain}\newtheorem{definition}[thm]{Definition} 

\begin{document}

\title[Secure Noisy Max with Gap]{A Floating-Point Secure Implementation of the Report Noisy Max with Gap Mechanism}

\author[Z.~Ding]{Zeyu Ding}	
\address{Binghamton University, Binghamton, NY 13902}	
\email{dding1@binghamton.edu}  

\author[J.~Durrell]{John Durrell}	
\address{Penn State University, University Park, PA 16802}	
\email{jmd6968@psu.edu}  

\author[D.~Kifer]{Daniel Kifer}	
\address{Penn State University, University Park, PA 16802}	
\email{dkifer@cse.psu.edu}  

\author[P.~Protivash]{Prottay Protivash}	
\address{Penn State University, University Park, PA 16802}	
\email{pxp945@psu.edu}  

\author[G.~Wang]{Guanhong Wang}	
\address{University of Maryland, College Park, MD 20742}	
\email{guanhong@umd.edu}  

\author[Y.~Wang]{Yuxin Wang}	
\address{Penn State University, University Park, PA 16802}	
\email{yxwang@psu.edu}  

\author[Y.~Xiao]{Yingtai Xiao}	
\address{Penn State University, University Park, PA 16802}	
\email{yxx5224@psu.edu}  

\author[D.~Zhang]{Danfeng Zhang}	
\address{Penn State University, University Park, PA 16802}	
\email{zhang@cse.psu.edu}  






\begin{abstract}
  \noindent 
  The Noisy Max mechanism and its variations are fundamental private selection algorithms that are used to select items from a set of candidates (such as the most common diseases in a population), while controlling the privacy leakage in the underlying data. A recently proposed extension, Noisy Top-k with Gap, provides numerical information about how much better the selected items are compared to the non-selected items (e.g., how much more common are the selected diseases). This extra information comes at no privacy cost but crucially relies on infinite precision for the privacy guarantees. In this paper, we provide a finite-precision secure implementation of this algorithm that takes advantage of integer arithmetic.
\end{abstract}

\maketitle

\section{Introduction}\label{sec:intro}
Differential privacy \cite{dp} is a de facto standard for data collectors to publish information about sensitive datasets while protecting the confidentiality of individuals who contribute data. It is widely adopted by government statistical agencies \cite{ashwin08:map,onthemap,Haney:2017:UCF,abowd18kdd,garfinkel18} and in industry \cite{rappor,BlockiDB16,prochlo,Ding17,appledp,elasticsensitivity,FBUrl}.
To achieve differential privacy, most algorithms introduce noise from  continuous probability distributions (e.g. the Laplace distribution) to mask the effect of any individual's data on the output of the algorithms. In practice, however, these distributions cannot be faithfully represented, much less sampled from, on computers which use only finite precision approximations (e.g., floating-point numbers) to real number arithmetic.

It might appear that such issues are purely of theoretical interest and do not cause serious harm in practice. Unfortunately, this is not the case: Mironov~\cite{fpvul} demonstrated that the textbook implementation of the Laplace Mechanism, the most basic algorithm to satisfy differential  privacy, can lead to catastrophic failures of privacy, causing entire datasets to be  reconstructed with a negligible privacy budget. This is due to a type of side channel attack that exploits the porous approximation of the reals using floating-point numbers. By examining the low-order bits of the noisy output, a large amount of infeasible candidate inputs can be eliminated and the true (noiseless) input value can often be determined. Furthermore, rounding the outputs of an insecure noise distribution does not resolve the problem~\cite{fpvul}. As a result of this demonstration, effort has been placed into developing  implementations of algorithms that \emph{exactly} sample from discrete distributions \cite{bv17,expmechbase2,permuteandflip,permuteflipnoisymax,Haney22}, given a source of uniform randomness. In many cases, these discrete distributions can replace the approximate (and insecure) continuous noise sampling algorithms that are found in standard statistical libraries.  For many differentially private mechanisms, appropriately rounding the inputs \cite{casacuberta2022widespread} and replacing their use of continuous noise with exact discrete samplers is enough to make their implementations secure from these floating point vulnerabilities. However, for the private selection mechanism we study in this paper, Noisy Top-k with Gap \cite{freegapinfo,freegapestimates}, such a drop-in replacement of discrete noise for continuous noise does not work and hence we propose a secure implementation.

Briefly, private selection mechanisms take a list of queries and a dataset as input. They output, with high probability, the identity of the query that has the largest value on the input dataset. For example, the \noisymax algorithm \cite{dpbook} takes a list of queries with sensitivity 1 (i.e., each query answer changes by at most 1 when a person is added to or removed from the data). It adds noise to each query answer and returns the identity of the query with the largest noisy answer.
Such mechanisms serve as key components in many privacy preserving algorithms for synthetic data generation \cite{mwem}, ordered statistics \cite{BlockiDB16}, quantiles \cite{smithconverge}, frequent itemset mining \cite{BhaskarDFP}, hyperparameter tuning \cite{LiuPSPC} for statistical models, etc. 
Recently, Ding et al. \cite{freegapinfo,freegapestimates} proposed novel variations of these selection mechanisms, including \noisymax \cite{dpbook}, that provide more functionality at the same privacy cost under pure differential privacy. 

In the case of the \noisymax algorithm, Ding et al. \cite{freegapinfo,freegapestimates} showed that, in addition to releasing the identity of the query whose noisy answer is the largest, it is possible to release a numerical estimate of the gap between the values of the  returned query and the next best query. This extra information comes at no additional cost to privacy, meaning that the original \noisymax mechanism threw away useful information. This result can be generalized to the setting in which one wants to estimate the identities of the top $k$ queries --  one can release (for free) estimates of all of the gaps between each top $k$ query and the next best query (i.e., the gap between the noisy best and noisy second best queries, the gap between the noisy second and noisy third best queries, etc). The generalized algorithm is called \gaptopk in their paper. For completeness, we include it as Algorithm~\ref{alg:gaptopk1} in Section~\ref{sec:alg}.

\gaptopk can release strictly more information at no additional cost to privacy which, when combined with subsequent noisy answers to each of the returned queries, can significantly reduce squared error of query estimates \cite{freegapinfo,freegapestimates}.
However, as explained in more detail in Section~\ref{subsec:hardness}, this algorithm is much more difficult to implement securely than \topk.
Its proof of privacy crucially relies on the differences (gaps) between pairs of random variables. Simply working in the integer domain and replacing Laplace with Discrete Laplace or exponential with the geometric distribution will not work as it invalidates the proof \cite{freegapinfo,freegapestimates} of privacy. So more involved alterations are needed to create a secure version of \gaptopk.

In this work, we propose an implementation of \gaptopk that is secure on finite computers. We make two modest assumptions. First,  the privacy loss budget parameter $\epsilon$ should be  a rational number. Second, the gaps should be represented as rational numbers and the user provides a desired denominator (integer) for these rational numbers. We refer to the reciprocal of this denominator as the \emph{target resolution $\tgtres$}, so that all returned gaps are multiples of $\tgtres$ (e.g., multiples of $2^{-10}$).

The secure algorithm is designed to be mathematically equivalent to running the ideal, infinite-precision algorithm, and then rounding all the gaps to the nearest multiple of $\tgtres$. Internally, the secure algorithm makes use of the exact geometric distribution sampler over a discrete domain \cite{dgauss}. It dynamically chooses the right level of finite precision it needs to work with so that its control flow matches what the ideal algorithm would have done. This is followed by a carefully controlled gap calculation step that is tricky because the gap (difference) between two discretized distributions is generally not the same as the discretized gap between two continuous distributions.

To summarize, our contributions are:
\begin{enumerate}
    \item We propose an implementation of the \gaptopk algorithm ~\cite{freegapinfo,freegapestimates} that is free from floating point vulnerabilities and can be implemented on finite precision machines.
    \item We prove the correctness of our implementation by carefully analysing its output distribution and show that it is equivalent to the original (ideal) \gaptopk (Algorithm~\ref{alg:gaptopk1}) followed by a post-processing step of rounding the gaps. 
    \item We evaluate our implementation on real and synthetic datasets and show that our secure implementation incurs moderate overhead over the (insecure) baseline algorithm.
\end{enumerate}

The rest of the paper is organized as follows. We discuss related work in Section~\ref{sec:related} and the relevant background in Section~\ref{sec:background}. We present our algorithms in Section~\ref{sec:alg} and proofs for their correctness in Section~\ref{sec:proofs}. We implement our algorithms, evaluate their performance on real datasets and report the results in Section~\ref{sec:experiments}. Finally we conclude in Section~\ref{sec:conclusion}.

\section{Related Work}\label{sec:related}
The floating point vulnerability in differentially private systems and its severity was first studied by Mironov~\cite{fpvul} and Gazeau et al. \cite{gazeau2016preserving}. They studied the effect of floating point on the resulting privacy guarantee. As an example, Mironov \cite{fpvul} demonstrated that by examining the low-order bits of the noisy outputs of the Laplace mechanism, the noiseless value can often be determined.
As a remedy, Mironov proposed the \emph{clamping mechanism} as a defence, but the defence tended to have worse utility than the Laplace mechanism.

A discrete alternative to the Laplace mechanism was proposed by Ghosh et al.~\cite{discretelaplace} and is equivalent to adding noise from a two-sided geometric distribution. The discrete Laplace mechanism satisfies $\epsilon$-differential privacy when the inputs are integers or appropriately rounded to integers and can be implemented exactly using the procedure proposed by  Cannone et al. \cite{dgauss}. In the same paper, Cannone et al. also showed how to exactly sample from a  discrete analogue of the Gaussian distribution~\cite{dgauss}, which can be used as a replacement for the continuous Gaussian when the inputs are integers or appropriately rounded. Holohan et al. \cite{Holohan21} proposed a method to sample Laplace and Gaussian noise in a way that makes it computationally harder for an attacker to reverse-engineer the output, without any rounding.

Secure implementations for several other mechanisms are also available, including  histogram approximation~\cite{bv17}, the Exponential Mechanism~\cite{expmechbase2} and the Noisy Max algorithm~\cite{permuteandflip,permuteflipnoisymax}. 
More recently, Haney et al. \cite{Haney22} proposed a sampling method called \emph{interval refining} which iteratively shrinks the internal $[0,1)$ until it is sufficiently small that the inverse images of both end points (hence all points in the interval) under the CDF of the sampling distribution round to the same floating point value. However, none of the above techniques directly apply to the \gaptopk algorithm as its privacy proof crucially depends on the difference between pairs of random variables~\cite{freegapinfo, freegapestimates}.  For example, if $Z_1, Z_2$ and $Z_3$ are random variables, existing techniques can correctly discretize each variable independently, but a secure implementation of \gaptopk may need to discretize pairwise differences like $Z_1-Z_2$ and $Z_2-Z_3$ -- these pairwise differences are clearly not independent of each other as they have variables in common.


\section{Background}\label{sec:background}
\subsection{Differential Privacy}
Differential privacy \cite {dp} is currently the gold standard for releasing privacy-preserving information about
a confidential database. It relies on the notion of adjacent datasets. Two datasets $D$ and $D'$ in some universe $\mathcal{D}$ are adjacent (or neighbors), denoted by $D\sim D'$, if they differ on the presence or absence of some  individual's data. Differential privacy ensures that the results of a computation on adjacent datasets are nearly indistinguishable. The degree of indistinguishability is quantified by a parameter $\epsilon>0$ called the privacy loss budget; the smaller $\epsilon$ is, the more privacy is provided.

\begin{defi}[Differential Privacy \cite{dp}]

Let $\epsilon > 0$. Let $\mech$ be randomized algorithm which takes a dataset $D\in \mathcal{D}$ as input.  Then $\mech$ satisfies (pure) $\epsilon$-differential privacy if for all pairs of adjacent datasets $D\sim D'\in \mathcal{D}\times \mathcal{D}$  and all output sets $S$, we have
\[\pr{\mech(D) \in S} \leq e^{\epsilon}\pr{\mech(D')\in S}\]
where the probability is over the randomness of the algorithm $\mech$.
\end{defi}

Differential privacy enjoys the following  properties:
\begin{itemize}
    \item Post-processing resilience. If the output of an $\epsilon$-differentially private algorithm $\mech$ goes through another computation $\mathcal{A}$ which does not use the datasets, then the composite algorithm $\mathcal{A}\circ \mech$ still satisfies $\epsilon$-differential privacy. In other words, privacy is not reduced by post-processing.
    
    \item Composition. If $\mech_1, \mech_2$ satisfy differential privacy with privacy loss budgets $\epsilon_1,\epsilon_2$, the algorithm that runs both and releases their outputs satisfies $(\epsilon_1+\epsilon_2)$-differential privacy. This result can be generalized to any finite number of differentially private algorithms.
\end{itemize}

Because of the compositional property of differential privacy, algorithms that satisfy differential privacy are usually built on smaller components called \emph{mechanisms}. Many differentially private algorithms take advantage of the Laplace mechanism \cite{dp}, which provides a noisy answer to a vector-valued function $f$ based on its $\ell_1$-\emph{global sensitivity} $\Delta_{f}$, defined as follows:

\begin{defi}[Global Sensitivity \cite{dpbook}]
The ($\ell_1$-)global sensitivity $\Delta_f$ of a vector-valued function $f$ with domain $\mathcal{D}$ is 
\[\Delta_f=\sup_{D\sim D'} \norm{f(D)-f(D')}_1\]
where the supremum is taken over all adjacent pairs $D\sim D'$ from $\mathcal{D}$.
\end{defi}

\begin{thm}[Laplace Mechanism \cite{dp}]\label{thm:laplace}
Given a privacy loss budget $\epsilon$, consider the mechanism that returns $\mech(D)=f(D)+ H$, where $H$ is a vector of independent random samples from the $\lap(\Delta_{f}/\epsilon)$ distribution.
Then $\mech$ satisfies $\epsilon$-differential privacy.
\end{thm}
Other kinds of additive noise distributions that can be used in place of Laplace in Theorem \ref{thm:laplace} include Discrete Laplace \cite{discretelaplace} and Staircase \cite{staircase}.

\subsection{Floating Point Vulnerability}

The most common method to sample a random variable $X$ from a distribution with a known cumulative distribution function (CDF) $F(\cdot)$ is the inverse sampling method: draw a sample $U$ from the uniform distribution on $[0,1)$ and apply the inverse CDF to obtain $X=F^{-1}(U)$. It's easy to check that the CDF of $X$ is indeed $F(\cdot)$:
$\forall x, P(X\leq x) = P(F^{-1}(U)\leq x) = P(U\leq F(x))=F(x).$
The inverse CDF of Laplace and exponential distributions are particularly simple. Thus most software libraries use this method to sample from these (and many other) distributions. 

However, the Laplace and exponential distributions are both continuous over the real numbers. As such, it is not possible to even represent a sample from them on a finite computer, much less to produce one. On one hand, given the non-uniform density of floating-point numbers, a uniform distribution over $[0,1)$ is not well-defined. On the other hand, floating-point operations involved in applying the inverse CDF will result in missing values and values that appear more frequently than they should~\cite{fpvul}.
It may seem that such issues are mainly of theoretical interest and do not cause serious harm in practice. Unfortunately, this is not the case: Mironov~\cite{fpvul} demonstrated that the textbook implementation of the Laplace Mechanism can lead to catastrophic failures of privacy. In particular, by examining the low-order bits of the noisy output, the noiseless value can often be determined. In one proof of concept attack, an entire dataset of 18K records was  reconstructed with a negligible ($<10^{-6}$) total privacy budget.

More recently, Casacuberta et al. \cite{Casacuberta22} showed that floating point calculations, unless carefully controlled, can cause finite machines to incorrectly compute the sensitivity of functions, thereby underestimating the amount of noise needed to protect privacy.

\subsection{Other Distributions}\label{sec:other}
In this paper, we also make use of the following distributions, whose notation we present here.

The exponential distribution with scale $\lambda/\epsilon$, written $\expo(\lambda/\epsilon)$, is a probability distribution over nonnegative real numbers and has probability density function $f(x)=\frac{\epsilon}{\lambda} e^{-\epsilon x/\lambda}$ and cumulative distribution function $F(x)=1-e^{-\epsilon x/\lambda}$.

The geometric distribution with success probability $p$, written $\geom(p)$, is a discrete distribution over the nonnegative integers $0, 1, 2, \dots$ with probability mass function $P(k) = (1-p)^kp$ and cumulative distribution function $F(k) = 1-(1-p)^{k+1}$.

Both distributions are memoryless in the sense that if $X$ is distributed either as an exponential or geometric random variable, then for any three nonnegative numbers $x,y,z$ such that $z>y$, we have $P(X\in [x+y, x+z] ~|~ X\geq x)=P(X \in [y, z])$.

These two distributions are also related by truncation. Let $\floor{x}$ denote the largest integer $\leq x$. Let $X\sim \expo(\lambda/\epsilon)$ and $Y\sim \geom(1-e^{-\epsilon/\lambda})$. Then comparing their CDFs, we see that for any nonnegative integer $k$,
$$P(X\leq k+1)=P(\floor{X}\leq k)=P(Y\leq k)$$

This means that a sample from an $\expo(\lambda/\epsilon)$ random variable $X$, which can be written as $\down{X}+(X-\down{X})$, is probabilistically equivalent to the sum of two \emph{independent} random variables $Y+X'$, where  $Y\sim \geom(1-e^{-\epsilon/\lambda})$, and $X'$ is a sample from $\expo(\lambda/\epsilon)$ conditioned on it being between $0$ and $1$.

\section{Algorithms}\label{sec:alg}
\subsection{The Noisy Top-k with Gap Algorithm}

The basic Noisy Max mechanism \cite{dpbook} and its generalization, Noisy Top-k, are fundamental private selection algorithms that are used to select items from a set of candidates (such as the most common diseases in a population), while controlling the privacy leakage in the underlying data. It takes a list of queries, each with sensitivity 1 (e.g., the count for any disease can change by at most 1 when a person is added to or removed from the data). It then adds noise to each query answer (e.g., the count for each disease) and returns the identity of the query with the $k$ largest noisy answers (e.g., the likely most common diseases). 

The preferred instantiation of the Noisy Max mechanism \cite{permuteandflip,permuteflipnoisymax} for pure $\epsilon$-differential privacy adds $\expo(2/\epsilon)$ noise, with pdf $f(x)=\frac{\epsilon}{2}e^{-\epsilon x/2}$, to each query answer and returns the identity (not value) of the query with the largest noisy answer. The extension to Noisy Top-k changes the noise parameter from $2/\epsilon$ to $2k/\epsilon$ and returns the identities of the queries with the top $k$ noisy answers.

A recently proposed extension, Noisy Top-k with Gap \cite{freegapinfo,freegapestimates}, provides numerical information about how much better the selected items are compared to the non-selected items. Its pseudo code is shown in Algorithm~\ref{alg:gaptopk1}. There are two major differences between this extension and the  Noisy Top-k mechanism: 1) The \gaptopk internally keeps track of the $(k+1)^\text{th}$ query (Line~\ref{line:gaptopkargmax}), and 2) The gaps (i.e., differences in values) between each query in the top $k$  and the corresponding next best query are calculated and returned together with the identities of the top $k$ queries  (i.e., the gap between the noisy best and noisy second best queries, the gap between the noisy second and noisy third best queries, etc., are returned), whereas the Noisy Top-k only returns query identities. Remarkably, this extra gap information comes at no additional privacy cost: both mechanisms satisfy pure differential privacy with the exact same privacy parameter $\epsilon$ \cite{freegapinfo}.

\begin{algorithm}[ht]
\SetKwProg{Fn}{function}{\string:}{}
\SetKwFunction{Test}{GapTopK}
\SetKwInOut{Input}{input}
\DontPrintSemicolon
\Fn{\Test{$q_1, \ldots, q_n, k, \epsilon$}}{
\For{$i= 1, \cdots, n$}{
$\exprv\gets \expo(2k/\epsilon)$\;
$\widetilde{q}_i \gets q_i+ \exprv$\;
}
$j_1, \ldots, j_{k+1} \gets \arg\max_{k+1}(\widetilde{q}_1,\ldots, \widetilde{q}_n)$~
\;\label{line:gaptopkargmax}
\For{$i = 1, \ldots, k$\label{line:gaptopkgap1}}
{
$g_i \gets \widetilde{q}_{j_i} - \widetilde{q}_{j_{i+1}}$\label{line:gaptopkgap2}\;
}
\Return $(j_1, g_1), \ldots, (j_k, g_k)$
\label{line:gaptopkret}
}
\caption{Noisy Top-k with Gap}
\label{alg:gaptopk1}
\end{algorithm}

Nevertheless, the extra gap information can offer better utility. For example, 
if an algorithm later asks for noisy answers to those queries that were returned by
\gaptopk, these noisy answers can be combined with the gaps to reduce their variance \cite{freegapinfo,freegapestimates}. Hence the gaps can be useful and that is why we focus on a secure implementation of this algorithm that is free of floating point side channels.

\subsection{Why the Gap Complicates the Implementation.}\label{subsec:hardness}
We next explain why \gaptopk is much more difficult to implement securely than Noisy Top-k. As we show next, the intuition is that without the gap, Noisy Top-k has categorical outputs (the ids of selected queries) and can be implemented by replacing the exponential distribution with the geometric distribution, taking advantage of the memorylessness properties of both. However, \gaptopk  adds continuous outputs and considers the \emph{pairwise} differences between noisy query answers. The pairwise differences of exponential or geometric random variables no longer have a memorylessness property to take advantage of. To see this more concretely, we sketch out the proof of  Noisy Max (without the gap) that only uses rational numbers and discrete probability distributions. Hence, it can be implemented without floating point vulnerabilities.

\begin{lem}\label{lem:snm}
Let $q_1,\dots,q_n$ be integer-valued queries, each with sensitivity 1. Let $\mech_{snm}$ be the mechanism such that, on input $D$, $\mech_{snm}(D)$ first computes $x_i\equiv q_i(D) + \georv_i$ where $\georv_i\sim\geom(1-e^{-\epsilon/2})$, then computes the set of maximums: $S=\{i~|~ x_i=\max(x_1,\dots, x_n)\}$, and returns a uniformly random sample from $S$. Then $\mech_{snm}$ satisfies $\epsilon$-differential privacy.
\end{lem}
\begin{proof}[Proof sketch]
The proof sketch is as follows. 
Consider the mechanism $\mech_{exp}$ that is identical to $\mech_{snm}$ except that $\mech_{exp}$ uses exponential noise instead of geometric. It first computes $x_i^\prime=q_i(D) + \exprv_i$, where $\exprv_i\sim \expo(2/\epsilon)$ and returns the $i$ for which $x_i^\prime$ is largest (because the noise distribution is continuous, ties occur with probability 0). Mechanism $\mech_{exp}$ is known to satisfy $\epsilon$-differential privacy \cite{permuteflipnoisymax}.

In terms of notation, $\mech_{snm}$ uses geometric random variables $\georv_i$ and $\mech_{exp}$ uses exponential random variables $\exprv_i$.
However, as explained in Section \ref{sec:other}, $q_1(D)+\exprv_1, \dots, q_n(D)+\exprv_n$ is  probabilistically equivalent to  $q_1(D)+\georv_1+\texprv_1, \dots, q_n(D)+\georv_n+\texprv_n$, where the $\texprv_i$ are all identically and independently distributed random variables with support $[0,1)$. This equivalence is true because the exponential and geometric distributions are memoryless.
Thus $\mech_{exp}$ can be thought of as computing the same set $S$ as $\mech_{snm}$, which is $S=\{i~:~q_i(D)+\georv_i=\max_j (q_j(D)+\georv_j)\}$ and then using the $\texprv_j$ to break ties. However, since the $\texprv_j$ are all i.i.d., using them to break ties is equivalent to picking an element of $S$ uniformly at random \cite{permuteflipnoisymax}. 

Thus $\mech_{snm}$ and $\mech_{exp}$ have the same exact probabilistic relationship between input and output, except that $\mech_{snm}$ only relies on discrete distributions.
\end{proof}

Note that Noisy Max (without the gap) provides only discrete outputs, so the conversion to using discrete noise was simple. However, a secure version of Noisy Max with Gap (i.e., \gaptopk with $k=1$) that uses discrete noise will need to provide a discretized gap. There are two seemingly intuitive ways of doing this:
\begin{itemize}
\item Option 1: Replace exponential noise in the algorithm (Algorithm \ref{alg:gaptopk1}) throughout with geometric noise. However, with geometric noise, there will often be ties  preventing the top $k+1$ from being unique. There will need to be a tie-breaking procedure and, unlike in Lemma \ref{lem:snm}, it is nontrivial because the distribution of the gaps depends on how many items are tied (i.e., it is related to order statistics). A direct adaptation of the original \gaptopk proof would result in a privacy parameter equal to $\epsilon/k$ times the number of noisy queries that are tied for a spot in the top $k$ (which is at least $k$ and is a random number).  Thus the provable privacy semantics would get worse.
\item Option 2: Create an algorithm that uses discrete noise, but is eqivalent to running the continuous version of  \gaptopk and discretizing (rounding) the gaps to the nearest multiple of some rational number $\tgtres$. This is the approach we take, but we explain the nontrivial difficulties it involves. Referring back to the notation in the proof sketch for Noisy Max, releasing a discretized gap would involve releasing a discretized version of  $q_i(D)+\exprv_i - (q_j(D)+\exprv_j)$, where $i$ is the query for which $q_i(D)+\exprv_i$ is largest and $j$ is the query for which $q_j(D)+\exprv_j$ is second largest. Even if we are lucky and there are no ties, the discretized versions of $q_i(D)+\exprv_i$ and  of $q_j(D)+\exprv_j$ do not provide enough information to compute the discretized version of their difference $q_i(D)+\exprv_i - (q_j(D)+\exprv_j)$.
The solution to this problem will require randomized rounding routines that rely on random permutations to determine how quantities are rounded. Secondly, breaking ties will require replacing the exponential distribution with a discrete distribution whose domain needs to be dynamically determined (e.g., instead of being over integers, it may need to be over multiples over some rational number that is not known in advance).
\end{itemize}

\subsection{Notation and Setup for the Secure Implementation.}
To describe the secure implementation, we use the following notation. 
Let $\tgtres$ denote the target resolution (a rational number), so that all returned gaps are multiples of $\tgtres$ (e.g., multiples of $2^{-10}$). In particular, $\tgtres$ is the reciprocal of an integer, so that all integer-valued query answers are multiples of $\tgtres$.
We use $\res$ for other resolutions that are refinements of $\tgtres$ (i.e., $\tgtres$ is always a multiple of $\res$).
For a positive integer $k$ we use $[k]$ to denote the set of integers from 1 to $k$: $[k]\triangleq\set{1, \ldots, k}$. We use $\pi$ denote a permutation on $[k]$, i.e., a bijective function $\pi:[k]\rightarrow [k]$. For a real number $x\in\bbR$, we use $\floor{x}$ to denote the floor of $x$ (the largest integer $\leq x$). We use $\down[\res]{x}$ to denote the largest multiple of $\res$ that is $\leq x$, which can be expressed mathematically as $\down[\res]{x}\triangleq\floor{\frac{x}{\res}}\cdot\res$. This notation is summarized in Table~\ref{tab:notation}.

\begin{table}[!ht]
\begin{center}
\caption{Notation}\label{tab:notation}
\begin{tabular}{c c}
\Xhline{1.5\arrayrulewidth}
\textbf{Symbol} & \textbf{Meaning} \\ \hline
$[k]$ &  $\set{1,\ldots,k}$  \\ 
$\pi$ & permutation on $[k]$ (bijective map $\pi: [k]\rightarrow [k]$) \\
$\tgtres$ & target resolution (e.g., $2^{-10}$), reciprocal of an integer\\
$\res$ & other resolutions that are refinements of $\tgtres$ (e.g., $2^{-11}, 2^{-12}, \ldots$)\\
$\floor{x}$ & floor of $x$ (the largest integer $\leq x$)\\
$\down[\res]{x}$ & rounding of $x$ down to the nearest multiple of $\res$, $\down[\res]{x}=\floor{\frac{x}{\res}}\cdot\res $ \\
\Xhline{1.5\arrayrulewidth}
\end{tabular}
\end{center}
\end{table}

The proofs and intermediate steps, that transform algorithms based on continuous noise into algorithms based on rational numbers and discrete noise, use distributions summarized in Table \ref{tab:randomnoise}. The two main distributions are the exponential and geometric distributions. For consistency we use $\exprv$ (resp. $\georv$) to denote a random variable following the exponential (resp. geometric) distribution. When one forms the corresponding conditional distributions, conditioned on the random variable being less than some threshold $\tau$, the result is a truncated distribution. In the case of exponential and geometric distributions, the truncated versions are the same as taking the original random variable modulo the threshold $\tau$.  We use $\texprv$ (resp. $\tgeorv$) to denote a random variable following the \emph{truncated} exponential (resp. \emph{truncated} geometric) distribution. We use $\sgeorv_{\res}$ to denote a ($\res$-) scaled geometric random variable, which is the same as a geometric random variable multiplied by $\gamma$, and so its domain is over integer multiples of $\res$.

\begingroup
\renewcommand*{\arraystretch}{1.5}
\begin{table}[!ht]
\begin{center}
\caption{Noise Distributions}\label{tab:randomnoise}
\begin{tabular}{c c c c }
\Xhline{1.5\arrayrulewidth}
\textbf{Distribution} & \textbf{Symbol} & \textbf{Support} & \textbf{Density/Mass} \\  \hline 
Exponential & $\exprv\sim\expo(\beta)$ & $[0,\infty)$ & $\frac{1}{\beta} e^{-\frac{x}{\beta}}$\\ 
Truncated exponential & $\texprv\sim\expo(\beta) \bmod \res$ & $[0,\res)$ & $\frac{1}{\beta} e^{-\frac{x}{\beta}}\big/(1-e^{-\frac{\res}{\beta}})$\\ 
Geometric & $\georv\sim\geom(p)$ & $\set{0, 1, 2,  \ldots}$ & $p(1-p)^{m}$ \\
Truncated geometric & $\tgeorv \sim\geom(p) \bmod M$ & $\set{0,1,\ldots,M\!-\!1}$ & $\frac{p(1-p)^{m}}{1-(1-p)^M} $\\
Scaled geometric & $\sgeorv_{\res}\sim\res\cdot\geom(p)$ & $\set{0, \res, 2\res, \ldots}$ & $ p(1-p)^{m}$ \\
\Xhline{1.5\arrayrulewidth}
\end{tabular}
\end{center}
\end{table}
\endgroup

\subsection{Secure Primitives for the Implementation.}

We use the following sampling primitives to implement our algorithms. These sampling algorithms do not use floating-point operations and are thus free of the floating-point vulnerability. The fundamental assumption is that there is access to a sequence of independent, uniformly random bits.

\paragraph*{\textbf{Random integer generation in an interval}} 
Given a sequence of uniformly random bits, one can generate random integers in a finite range, e.g. $0,1,\dots, k-1$,
using rejection sampling~\cite{vonNeumann1951,gnucpp,golang,lemirerandom}.

\paragraph*{\textbf{The Fisher-Yates random shuffle}}
Another commonly used technique is the Fisher-Yates shuffle~\cite{fisheryates,Knuth69}. It randomly permutes a list of $n$ elements in place so that all $n!$ possible permutations are equally probable. This algorithm is based on uniform sampling of integers.

\paragraph*{\textbf{Sampling from a Bernoulli distribution}}
For certain values of success probability $p$ (even irrational values), it is possible to reduce sampling from $\bern(p)$ to sampling random integers without any use of floating point. First, when $p=n/d$ is a rational number, 
it suffices to draw an integer $N$ uniformly from the range $[0,d)$ and output 1 if $N<n$ and 0 otherwise. When $p=e^{-\rat}$ for some positive rational number $\rat\in\bbQ^+$, Cannonne et al.~\cite{dgauss,Canonne_Kamath_Steinke_2022} reduced the task of sampling from $\bern(e^{-\rat})$ to that of sampling from $\bern(\rat/k)$ for various integers $k\geq1$ (see Algorithm 1 in \cite{Canonne_Kamath_Steinke_2022}), without the need for computing $e^{-\rat}$.

\paragraph*{\textbf{Sampling from a geometric distribution}} Canonne et al.~\cite{dgauss,Canonne_Kamath_Steinke_2022} also showed that for certain values of success probability $p$, namely those in the form of $1-e^{-\rat}$ where $\rat$ is a positive rational number, sampling from a $\geom(1-e^{-\rat})$ distribution can be efficiently reduced to sampling from $\bern(e^{-\rat^\prime})$, where $\rat^\prime$ is a rational number that depends on $\rat$. Canonne et. al sample a geometric inside of a discrete Laplace sampling algorithm (Algorithm 2 in \cite{Canonne_Kamath_Steinke_2022}), so for completeness, we extract the geometric sampler from their algorithm and list it in the appendix.
Like before, this algorithm completely avoids use of floating point.

\subsection{A Secure Implementation of Noisy Top-k with Gap}

A common approach to making a differentially private algorithm secure on finite machines is to take the ideal algorithm $\mathcal{M}$ (Algorithm~\ref{alg:gaptopk1} in our case) that assumes infinite precision and define an intermediate ``rounded'' version $\mathcal{M}'$ that differs from $\mathcal{M}$ in only two ways. First, the inputs to $\mathcal{M}'$ are rounded to rational numbers to avoid floating point inputs. Then $\mathcal{M}'$ rounds the output of $\mathcal{M}$ to rational numbers. The final step is to create a secure algorithm $\mathcal{M}''$ that completely avoids floating point in its intermediate state while still being probabilistically equivalent to $\mathcal{M}'$ -- in other words, for every dataset $\data$, the output distributions $P(\mathcal{M}'(\data))$ and $P(\mathcal{M}''(\data))$ are the same. 
Following this approach, we first create a rounded-input/output version of Algorithm~\ref{alg:gaptopk1} by rounding the inputs and each gap in the output down to its nearest multiple of $\tgtres$. 
Thus $\mathcal{M}'$ from the above discussion is  the following Algorithm~\ref{alg:gaptopk2} (note that the input rounding happens on Line \ref{line:gaptopk2round}).
Then, to create the final secure algorithm $\mathcal{M}''$, we proceed with two steps. First, we create another intermediate algorithm (Algorithm~\ref{alg:gaptopk3}) that has both continuous random variables and discrete random variables that are derived from the continuous ones. Last we create $\mathcal{M}''$ (Algorithm~\ref{alg:gaptopk4}) by removing the continuous random variables, while keeping the joint distribution of the discrete random variables unchanged.

\begin{algorithm}[ht]
\SetKwProg{Fn}{function}{\string:}{}
\SetKwFunction{Test}{GapTopK}
\SetKwInOut{Input}{input}
\Input{$q_1,\dots, q_n$:   answers to sensitivity 1 queries.\\$\tgtres$ is the target resolution; 1 is a multiple of $\tgtres$\\$\epsilon$ is rational}
\DontPrintSemicolon
\Fn{\Test{$q_1, \ldots, q_n, k, \epsilon$}}{
\For{$i= 1, \cdots, n$}{
$\exprv\gets \expo(2k/\epsilon)$\;\label{line:gaptopk2noise}~
$\widetilde{q}_i \gets \down[\tgtres]{q_i}+ \exprv$\;\label{line:gaptopk2round}
}
$j_1, \ldots, j_{k+1} \gets \arg\max_{k+1}(\widetilde{q}_1,\ldots, \widetilde{q}_n)$\;\label{line:alg2_index}
\For{$i = 1, \ldots, k$}{
$g_i \gets \down[\tgtres]{\widetilde{q}_{j_i} - \widetilde{q}_{j_{i+1}}}$\;
}
\Return $(j_1, g_1), \ldots, (j_k, g_k)$
}
\caption{Noisy Top-k with Gap (Rounded)}
\label{alg:gaptopk2}
\end{algorithm}

To show that Algorithm~\ref{alg:gaptopk2} is differentially private, we first show that rounding down does not affect the sensitivity.
\begin{lem}\label{lem:rounddown_sensitivity}
Suppose $q$ is a scalar function with $\ell_1$-sensitivity $\Delta$. Suppose that $\res$ is a positive number and that $\Delta$ is an integer multiple of $\res$. Then $\down[\res]{q}$ also has $\ell_1$-sensitivity $\Delta$.
\end{lem}
\begin{proof}
Let $D$ and $D'$ be two neighboring datasets that differ on one person.
Without loss of generality, assume $q(D)\geq q(D^\prime)$. Then 
$0 \leq q(D)-q(D^\prime) \leq \Delta$ which means there exists a nonnegative integer $k$ with $k\leq \Delta/\res$  and a number $s\in[0,\res)$ such that $q(D)-q(D^\prime)+ k\gamma + s=\Delta$
and so
$q(D) - \Delta + k\res +s = q(D^\prime)$.
Let $\remainder(q(D)) = q(D) -\down[\res]{q(D)}$. Then 
$$\down[\res]{q(D)} - \Delta + k\res  + \remainder(q(D)) + s = \down[\res]{q(D^\prime)} + \remainder(q(D^\prime))$$
Noting that rem$(q(D))+s \in [0, 2\res)$, this means
$$\down[\res]{q(D)} - \Delta + k^\prime\res  + s^\prime = \down[\res]{q(D^\prime)} + \remainder(q(D^\prime))$$
where $k^\prime\in\{k, k+1\}$ and $s^\prime \in [0,\res)$. This means that $s^\prime=\remainder(q(D^\prime))$ as both quantities are nonnegative and less than $\res$, while everything else is a nonnegative multiple of $\res$. Rounding both sides down to the nearest multiple of $\res$, we get 
$$\down[\res]{q(D)} - \Delta + k^\prime\res  = \down[\res]{q(D^\prime)}$$
where $k^\prime$ is an integer between $0$ and $\Delta/\res+1$. Thus
$$
    \down[\res]{q(D)} - \down[\res]{q(D^\prime)} = \Delta - k^\prime\res \in [-\res, \Delta]\subseteq [-\Delta, \Delta]
$$
since $\Delta$ is a multiple of $\res$.
\end{proof}

\begin{thm}
Algorithm~\ref{alg:gaptopk2} is $\epsilon$-differentially private.
\end{thm}
\begin{proof}
Algorithm~\ref{alg:gaptopk2} is simply Algorithm~\ref{alg:gaptopk1} applied to sensitivity 1 queries $\down[\tgtres]{q_i(D)}$ followed by a post-processing step (which therefore does not affect privacy parameters) of rounding every gap down to the closest multiple of $\tgtres$.
\end{proof}

Thus, from now on, we assume that the query answers are integer multiples of $\tgtres$.
Next, to obtain a secure implementation of Algorithm~\ref{alg:gaptopk2}, we need to create an algorithm whose internal states are machine representable (i.e., rational numbers), and whose output distribution is identical to that of Algorithm~\ref{alg:gaptopk2}. This is challenging because as pointed out previously, any straightforward discretization of the noisy query answers $\widetilde{q}_i$ (e.g., $\down[\tgtres]{\widetilde{q}_i}$) will cause the probability of ties to be nonzero. Furthermore, it can happen that $\down[\tgtres]{\widetilde{q}_i-\widetilde{q}_j}\neq \down[\tgtres]{\widetilde{q}_i} - \down[\tgtres]{\widetilde{q}_j}$ so the discretiazed gaps are not readily available from the discretized noisy queries. We solve this problem by a  two-step approach. The approach is described here and proofs are presented in Section \ref{sec:proofs}.

\begin{algorithm}[ht]
\SetKwProg{Fn}{function}{\string:}{}
\SetKwFunction{Test}{GapTopK}
\SetKwInOut{Input}{input}
\DontPrintSemicolon
\Input{$\tgtres$ is the target resolution; 1 is a multiple of $\tgtres$\\$q_1,\dots, q_n$:  answers to sensitivity 1 queries,  all are integer multiples of $\tgtres$.\\$\epsilon$ is rational\\$M$ is an integer used to refine the resolution.}
\Fn{\Test{$q_1, \ldots, q_n, k, \epsilon$}}{
\For{$i= 1, \cdots, n$}{
$\exprv\gets \expo(2k/\epsilon)$\; \label{line:gaptopk3noise}~ 
$\widetilde{q}_i \gets q_i+ \exprv$, \quad $\widehat{q}_i^{\,(0)} \gets \down[\tgtres]{\widetilde{q}_i}$\;\label{line:gaptopk3disc}
}
$j_1, \ldots, j_{k+2} \gets \arg\max_{k+2}(\widehat{q}_{1}^{\,(0)},\ldots, \widehat{q}_{n}^{\,(0)})$\;\label{line:gaptop3firstmax}

$t \gets 0$, \quad $\res_0 \gets \tgtres$\label{line:alg3_index_start}\;
\While{\textrm{there is a tie among \label{line:gaptopk3while}} $\widehat{q}_{j_1}^{\,(t)}, \ldots, \widehat{q}_{j_{k+2}}^{\,(t)}$}{
$t \gets t+1$, \quad $\res_t\gets \frac{1}{M}\res_{t-1}$\;\label{line:alg3gamma}
\For{$i=1, \ldots, n$}
{
$\widehat{q}_{i}^{\,(t)} \gets \down[{\res_t}]{\widetilde{q}_i}$\label{line:alg3qhat}
}
$j_1, \ldots, j_{k+2} \gets \arg\max_{k+2}(\widehat{q}_{1}^{\,(t)},\ldots, \widehat{q}_{n}^{\,(t)})$\;
}
\For{$i = 1, \ldots, k$}{
$g_i \gets \down[\tgtres]{\widetilde{q}_{j_i} - \widetilde{q}_{j_{i+1}}}$\;\label{line:alg3gap}
}
\Return $(j_1, g_1), \ldots, (j_k, g_k)$
}
\caption{Noisy Top-k with Gap (Intermediate)}
\label{alg:gaptopk3}
\end{algorithm}

The first step is to create an intermediate algorithm (Algorithm \ref{alg:gaptopk3}) that has both continuous random variables and discrete random variables that are derived from them. The last step (Algorithm \ref{alg:gaptopk4}) will remove the continuous random variables, while keeping the joint distribution of the discrete random variables unchanged.

In the intermediate Algorithm \ref{alg:gaptopk3},  discrete variables $\widehat{q}_i^{\,(0)}$ are initially introduced by  discretizing the continuous noisy query answers $\widetilde{q}_i$ (Line \ref{line:gaptopk3disc}). These discrete variables are used for the decision-making (i.e., determining the winning queries) in Line \ref{line:gaptop3firstmax}. 
Because of the possibility of ties in discrete random variables, 
 several new features are introduced by Algorithm~\ref{alg:gaptopk3}. 
\begin{itemize}
\item First, we get the top $k+2$ noisy query answers instead of $k+1$ answers. This is done so that we know if there are ties in the $k+1^\text{th}$ place. If there are ties anywhere in the top $k+1$ places, we will need to break them using the second feature of Algorithm \ref{alg:gaptopk3}, described next. 

\item  Next, we introduce a tie-breaking loop (Line \ref{line:gaptopk3while}) which keeps going until all ties in the top $k+1$ places are resolved.  If there is a tie between any two  such queries, the algorithm increases the precision of rounding -- instead of rounding the continuous noisy answers down to a multiple of the target resolution $\tgtres$, the algorithm revisits those noisy answers and rounds them down to a multiple of $\tgtres/M$ (where the integer $M>1$ is an algorithm parameter that is 10 in our experiments). This results in the next version of the discrete random variables $\widehat{q}_i^{\,(1)}$. This finer precision may break the ties, and if not, the precision is refined again until eventually all ties are broken. Later, Algorithm \ref{alg:gaptopk4} will simulate this loop using only discrete random variables.  
\end{itemize}
After the distinct top $k+1$ queries are determined, the gaps are computed. However, for now, the discretized gaps are still computed by rounding the gaps computed from continuous noisy answers (this problem will be solved by Algorithm \ref{alg:gaptopk4}).

\begin{algorithm}[ht]
\SetKwProg{Fn}{function}{\string:}{}
\SetKwFunction{Test}{GapTopK}
\SetKwInOut{Input}{input}
\DontPrintSemicolon
\Input{$\tgtres$ is the target resolution; 1 is a multiple of $\tgtres$\\$q_1,\dots, q_n$:  answers to sensitivity 1 queries,  all are integer multiples of $\tgtres$.\\$\epsilon$ is rational\\$M$ is an integer used to refine the resolution.}
\Fn{\Test{$q_1, \ldots, q_n, k, \epsilon$}}{
\For{$i= 1, \cdots, n$}{ \label{line:part1begin}
$\georv_0\gets \geom(1-e^{-\epsilon\tgtres/2k})$\label{line:gaptopk4noise}\; 
$\widecheck{q}_{i}^{\,(0)} \gets q_i+\tgtres\cdot\georv_0$\;\label{line:gaptopk4disc}
}
$j_1, \ldots, j_{k+2} \gets \arg\max_{k+2}(\widecheck{q}_{0}^{\,(0)},\ldots, \widecheck{q}_{n}^{\,(0)})$\; \label{line:part1end}
$t\gets 0$, \quad $\res_0 \gets \tgtres$\; \label{line:part2begin}
\While{there is a tie among $\widecheck{q}_{j_1}^{\,(t)}, \ldots, \widecheck{q}_{j_{k+2}}^{\,(t)}$ }{
$t\gets t+1$, \quad $\res_t \gets \frac{1}{M}\res_{t-1}$\;
\For{$i=1, \ldots, n$}
{
$\georv_t\gets \geom(1-e^{-\epsilon\res_{t}/2k})$\label{line:gaptopk4noise2}\;  $\widecheck{q}_{i}^{\,(t)} \gets \widecheck{q}_{i}^{\,(t-1)}+\res_t\cdot(\georv_t \bmod M)$\label{line:gaptopk4noise2b}\;
}
$j_1, \ldots, j_{k+2} \gets \arg\max_{k+2}(\widecheck{q}_{1}^{\,(t)},\ldots, \widecheck{q}_{n}^{\,(t)})$\; \label{line:part2end}
}
%
$x_1,\ldots,x_{k+1}\gets \texttt{Shuffle}(1,\ldots,k+1)$~\tcp*[r]{Fisher-Yates shuffle}\label{line:part3begin}
\For{$i = 1, \ldots, k$ \label{line:gaptopk4gapstart}} 
{ 
\uIf{$x_i<x_{i+1}$}
{
$g_i \gets \down[\tgtres]{\widecheck{q}_{j_i}^{\,(t)} - \widecheck{q}_{j_{i+1}}^{\,(t)} - \res_t}$\;\label{line:alg4gap1}
}
\Else{
$g_i \gets \down[\tgtres]{\widecheck{q}_{j_i}^{\,(t)} - \widecheck{q}_{j_{i+1}}^{\,(t)} }$\;\label{line:alg4gap2}\label{line:part3end}
}
}\label{line:gaptopk4gapend}
%
\Return $(j_1, g_1), \ldots, (j_k, g_k)$
}
\caption{Noisy Top-k with Gap (Secure)}
\label{alg:gaptopk4}
\end{algorithm}

The next algorithm, Algorithm \ref{alg:gaptopk4} switches to only using discrete random variables, so that 
 all internal states are rational numbers. This involves removing the exponential random variables that were used by Algorithm \ref{alg:gaptopk3}, simulating the tie-breaking loop using only discrete random variables, and finally simulating the gap using only the discrete random variables. We handle the difficulties as follows:
 \begin{itemize}
 \item Since the continuous random variable $\widetilde{q}_i$ (query answer plus exponential noise) is rounded down to a multiple of $\tgtres$ to get $\widehat{q}_i^{\,(0)}$, the distribution of the latter is the same as the query answer plus geometric noise over multiples of $\tgtres$ (see Section \ref{sec:other}). These are represented as the random variables $\widecheck{q}_i^{\,(0)}$ in Line \ref{line:gaptopk4disc} of Algorithm \ref{alg:gaptopk4}.
 \item The discrete variable $\widehat{q}_i^{\,(j)}=\down[\tgtres/M^j]{\widetilde{q}_i}$ is equal to the continuous variable $\widetilde{q}_i$ rounded down to a multiple of $\tgtres/M^j$. But this also means that $\widehat{q}_i^{\,(j)}$ is equal to $\widehat{q}_i^{\,(j+1)}$ rounded down to a multiple of $\tgtres/M^j$ (i.e., $\widehat{q}_i^{\,(j)}=\down[\tgtres/M^j]{\widehat{q}^{\,(j+1)}_i}$). This allows us to use a modified version of the memoryless property of geometric distributions to simulate $\widehat{q}_i^{\,(j+1)}$ from $\widehat{q}_i^{\,(j)}$ and their joint distribution would be the same as if they were derived from the continuous random variable. Namely the conditional distribution of the part that was rounded down, which is $\widehat{q}^{\,(j+1)}_i-\widehat{q}^{\,(j)}_i = \widehat{q}^{\,(j+1)}_i - \down[\tgtres/M^j]{\widehat{q}^{\,(j+1)}_i}$, conditioned on $\widehat{q}^{\,(j)}_i$,  is a truncated geometric in the interval $[0, \tgtres/M^j)$ with support equal to integer multiples of $\tgtres/M^{j+1}$. In Algorithm \ref{alg:gaptopk4}, these variables are called $\widecheck{q}_i^{\,(j+1)}$ and $\widecheck{q}_i^{\,(j)}$ and the generation of the former from the latter using truncated geometric noise is performed in Lines  
 \ref{line:gaptopk4noise2} and \ref{line:gaptopk4noise2b}.
 \item Next we consider the intuition behind how Algorithm \ref{alg:gaptopk4} simulates the discretized gap from these discrete random variables. In general, for any two numbers $\exprv_i, \exprv_j\in\bbR$ and resolution $\res$, the quantities  $\down[\res]{\exprv_i-\exprv_j}$ and $\down[\res]{\exprv_i}-\down[\res]{\exprv_j}$ are not equal, but they differ by a correction term that we can simulate. Specifically, we use this fact from Lemma \ref{roundinglem} in Section \ref{sec:proofs}.\[
\down[\res]{\exprv_i-\exprv_j}=\down[\res]{\exprv_i}-\down[\res]{\exprv_j} -\delta_{ij}\res \quad \textrm{ where } 
\delta_{ij}=
\begin{cases}
0 &\quad \textrm{if } \exprv_i-\down[\res]{\exprv_i}\geq \exprv_j-\down[\res]{\exprv_j}\\
1 & \quad \textrm{otherwise }\\
\end{cases}
\]
We use this fact as follows. In the notation of intermediate Algorithm \ref{alg:gaptopk3}, $\exprv_i$ is the same as the continuous noisy answer $\widetilde{q}_i$. Given a resolution $\res$, the values $\widetilde{q}_i-\down[\res]{\widetilde{q}_i}$ are, due to the memoryless property, truncated exponentials supported on $[0,\res)$ and they are i.i.d. for all $i$. Therefore, the ordering of the $\widetilde{q}_i-\down[\res]{\widetilde{q}_i}$ values, for all $i$, is uniformly random. Hence, Line \ref{line:part3begin} in Algorithm \ref{alg:gaptopk4} determines this ordering with a random permutation and using that it determines how to simulate the discretized gap correction term.
 \end{itemize}
Now note that Algorithm \ref{alg:gaptopk4} only uses rational numbers ($1/\tgtres$ is an integer and $\epsilon$ must be rational) and discrete distributions (the parameter of the geometric distribution is not explicitly computed). The randomness primitives are:
\begin{itemize}
\item Generating an exact sample from a geometric distribution with parameter $(1-e^{-\epsilon\tgtres/2k})$ can be done using the algorithm from Canonne et al. \cite{Canonne_Kamath_Steinke_2022} (reproduced in the appendix here) without ever computing $e^{-\epsilon\tgtres/2k}$.
\item Generating a truncated geometric distribution can be performed by generating a geometric distribution and computing the result modulo a number as in Line \ref{line:gaptopk4noise2b}.
\item A random permutation can be generated from the Fischer-Yates shuffle algorithm \cite{fisheryates} and only requires choosing a random integer from a bounded range, which can also be performed exactly given a source of uniform bits \cite{Canonne_Kamath_Steinke_2022}.
\end{itemize}

Algorithm~\ref{alg:gaptopk4} can be further improved for efficiency. Note that if $\down[\res_t]{X_i} < \down[\res_t]{X_j}$, then for any refinement $\res_{t'}<\res_t$ ($\res_t$ is a multiple of $\res_{t'}$), we also have $\down[\res_{t'}]{X_i} < \down[\res_{t'}]{X_j}$. Thus if at some resolution $\res_t$, it can be determined that a query $q_i$ is not among the top candidates that contribute to the output indexes and gaps, then this query can be dropped. Thus we can maintain a pool of queries that are relevant to the output of the algorithm (i.e., those whose noisy answers at  the current resolution are tied with something in the top $k+1$) and only refine the noise to those queries. Initially, the pool will have all $n$  queries. But after the first iteration, the pool will have only roughly $O(k)$ queries left, making subsequent iterations much faster if $k\ll n$. We call this optimization \emph{early query pruning} and report its performance in Section~\ref{sec:experiments}.

\section{Proofs}\label{sec:proofs}
In this section, we prove the privacy guarantees of Algorithm~\ref{alg:gaptopk3} and Algorithm~\ref{alg:gaptopk4} by showing that they are equivalent to Algorithm~\ref{alg:gaptopk2}.

\begin{definition}
We say that two algorithms $\mech_1$ and $\mech_2$ are equivalent if for all datasets $D$ and all measurable sets $S$, $P(\mech_1(D)\in S)=P(\mech_2(D)\in S)$.
\end{definition}

\begin{thm}\label{thm:alg3}
Algorithm~\ref{alg:gaptopk3} is equivalent to Algorithm~\ref{alg:gaptopk2} and therefore is $\epsilon$-differentially private.
\end{thm}

\begin{proof}[Proof of Theorem~\ref{thm:alg3}]
First, we show that Algorithm~\ref{alg:gaptopk3} terminates with probability 1. Since $\widetilde{q}_1,\ldots,\widetilde{q}_n$ uses continuous noisy, with probability 1 there is no tie among $\widetilde{q}_1,\ldots,\widetilde{q}_n$. Therefore, let $\delta=\min_{i\neq j}\abs{\widetilde{q}_i-\widetilde{q}_j}$, then we have $\delta>0$ with probability 1. Thus when $\res_t=\tgtres/M^t\leq\delta$, i.e. $t\geq \log_M(\frac{\tgtres}{\delta})$, we have no ties among $\widehat{q}_1^{\,(t)},\ldots, \widehat{q}_n^{\,(t)}$. 

Next, we show that when the loop terminates, the indices $j_1,\ldots,j_{k+1}$ are indeed from the top $k+1$ noisy queries. 
Note that $\down[\res]{x}> \down[\res]{y}\implies x>y$.
By the termination condition, at the end of the loop we have that $\widehat{q}_{j_1}>\ldots>\widehat{q}_{j_{k+1}}>\widehat{q}_{j_{k+2}}\geq \widehat{q}_{s}, s\notin \set{j_1,\ldots,j_{k+2}}$. Therefore, we have $\widetilde{q}_{j_1}>\ldots>\widetilde{q}_{j_{k+1}}> \widetilde{q}_{s}, s\notin \set{j_1,\ldots,j_{k+1}}$. Thus $j_1,\ldots,j_{k+1}$ are the indices of the largest $k+1$ queries among $\widetilde{q}_1,\ldots,\widetilde{q}_n$.
This means if $\widetilde{q}_1,\ldots,\widetilde{q}_n$ are the same in Algorithm~\ref{alg:gaptopk2} and Algorithms~\ref{alg:gaptopk3}, then the indices $j_1,\ldots,j_{k+1}$ are the same. Since Algorithm~\ref{alg:gaptopk2} and Algorithm~\ref{alg:gaptopk3} only differ in how they choose the top $k+1$ noisy queries and the winning queries are the same, the two algorithms are equivalent.
\end{proof}

\begin{thm}\label{thm:alg4}
Algorithm~\ref{alg:gaptopk4} is equivalent to Algorithm~\ref{alg:gaptopk3} and therefore is $\epsilon$-differentially private. 
\end{thm}

This proof contains multiple stages and requires several supporting results.
To establish the equivalence between Algorithm~\ref{alg:gaptopk3} and Algorithm~\ref{alg:gaptopk4}, we need a few lemmas to establish the connection between successive roundings of an exponential random variables (with increasing resolutions) and a sequence of independent geometric random variables.

\begin{lem}\label{lem:exp_rounddown}
Let $\exprv\sim\expo(\beta)$. Then the distribution of $\sgeorv_{\res} \triangleq\down[\res]{\exprv}$ is the scaled geometric distribution over $\set{0,\res ,2\res,\dots}$ with success probability $p=1-e^{-\frac{\res}{\beta}}$. In other words, $\down[\res]{\exprv}$ follows the same distribution as $\res\cdot\georv$ where $\georv$ geometric distribution (over $\set{0,1,2,...}$) with success probability $p=1-e^{-\frac{\res}{\beta}}$.
\end{lem}
\begin{proof}
For any value $m\in\set{0,1,\ldots}$ we have
\begin{align*}
    P(\down[\res]{\exprv}=m\res) &= P(m\res \leq \exprv < m\res + \res)
           = \int_{m\res}^{m\res+\res} \frac{1}{\beta} e^{- \frac{x}{\beta}}~dx
           = -e^{-\frac{x}{\beta}}\Big|_{m\res }^{m\res+\res}\\
           &= e^{-\frac{ m\res}{\beta} } - e^{-\frac{ m\res+\res}{\beta}}
           = e^{-\frac{m\res}{\beta}} (1-e^{-\frac{\res}{\beta}}) =(1-p)^mp
\end{align*}
where we let $p=1-e^{-\frac{\res}{\beta}}$ and hence $(1-p)^m=e^{-\frac{m\res}{\beta}}$.
\end{proof}

\begin{lem}\label{lem:reverse_exp_rounddown}
Let $\exprv\sim\expo(\beta)$ and  $\texprv=\exprv-\down[\res]{\exprv}$. Then $\texprv$ follows the truncated exponential distribution on $[0,\res)$.
Furthermore, $P(\texprv\mid \down[\res]{\exprv})=P(\texprv)$, i.e., $\texprv$ and $\down[\res]{\exprv}$ are independent.
\end{lem}
\begin{proof}
By definition of $\down[\res]{\exprv}$ we have $\texprv\in[0,\res)$. For any $x\in [0,\res)$, 
\[\texprv\leq x\implies \exprv-\down[\res]{\exprv} \leq x\implies \exprv\in[l\res,l\res+x], l=0,1,\ldots.\]
Thus
\begin{align*}
    P(\texprv\leq x) &= \sum_{l=0}^\infty P(l\res \leq \exprv < l\res + x)
           = \sum_{l=0}^\infty\int_{l\res}^{l\res+x} \frac{1}{\beta} e^{-\frac{s}{\beta}}~ds\\
           &= \sum_{l=0}^\infty e^{-\frac{l\res}{\beta}} (1-e^{-\frac{x}{\beta}})
           =(1-e^{-\frac{x}{\beta}})\sum_{l=0}^\infty e^{-\frac{l\res}{\beta}} 
           =\frac{1-e^{-\frac{x}{\beta}}}{1-e^{-\frac{\res}{\beta}}} 
\end{align*}
Hence the density
\[
f(\texprv = x) = \frac{d}{dx} P(\texprv\leq x) =\frac{\frac{1}{\beta} e^{-\frac{x}{\beta}}}{1-e^{-\frac{\res}{\beta}}} \cdot \one{[0, \res)}.
\]
The independence of $\texprv$ and $\down[\res]{\exprv}$ is due to the memoryless property of exponential distribution:
\begin{align*}
   P(\texprv \leq x \mid \down[\res]{\exprv}=l\res) &=\frac{P(l\res\leq \exprv \leq l\res+x)}{P(l\res\leq \exprv \leq l\res+\res)}=\frac{e^{-\frac{l\res}{\beta}}-e^{-\frac{l\res+x}{\beta}}}{e^{-\frac{l\res}{\beta}}-e^{-\frac{l\res+\res}{\beta}}} = \frac{1-e^{-\frac{x}{\beta}}}{1-e^{-\frac{\res}{\beta}}} = P(\texprv\leq x).
\end{align*}
\end{proof}

\begin{lem}\label{lem:trunc_geo}
Let $\georv\sim \geom(p)$ and $\tgeorv= \georv \bmod M$ for some $M>1$. Then the distribution of $\tgeorv$ is the truncated geometric distribution with success probability $p$ on $\set{0, \ldots, M-1}$.
\end{lem}
\begin{proof}
For any value $m\in\set{0,\ldots, M-1}$
\begin{align*}
    P(\tgeorv = m) &= \sum_{l=0}^\infty P(\georv = lM+m)
          = \sum_{l=0}^\infty (1-p)^{lM+m}p
          = (1-p)^{m}p\sum_{l=0}^\infty (1-p)^{lM}\\
          &=\frac{(1-p)^{m}p}{1-(1-p)^M} 
\end{align*}
\end{proof}

\begin{lem}\label{lem:exp_rounddown_refine} Let $M$ be a positive integer. Let $\res_1, \res_2>0$ be such that $\res_1=M\res_2$. Let $\exprv\sim\expo(\beta)$, $\sgeorv_{\res_1}=\down[\res_1]{\exprv}$ and $\sgeorv_{\res_2}=\down[\res_2]{\exprv}$. Then $\sgeorv_{\res_1}=\down[\res_1]{\sgeorv_{\res_2}}$ and $P(\sgeorv_{\res_2}~|~\sgeorv_{\res_1})$ is the same as the probability mass function of $\sgeorv_{\res_1} + \res_2(\georv \bmod M)$ where $\georv\sim\geom(1-e^{-\frac{\res_2}{\beta}})$ is independent of $\sgeorv_{\res_1}$.
\end{lem}

\begin{proof}
First we show $\sgeorv_{\res_1}=\down[\res_1]{\sgeorv_{\res_2}}$. Intuitively, this means that the effect of rounding down a number to a finer resolution ($\res_2$) first then rounding down the result again to a coarser resolution ($\res_1$) is the same as rounding the number directly to the coarser resolution ($\res_1$).
Let $\sgeorv_{\res_1}=n\res_1$, then $\exprv = n\res_1 + s$ for some $s\in[0,\res_1)$. Thus we have $\sgeorv_{\res_2}=\down[\res_2]{\exprv}=n\res_1+\down[\res_2]{s}$ because $n\res_1$ is already a multiple of $\res_2$. Since  $\down[\res_2]{s}\leq s<\res_1$, we have $\down[\res_1]{\down[\res_2]{s}}=0$ and $\down[\res_1]{\sgeorv_{\res_2}}=n\res_1=\sgeorv_{\res_1}$.
Moreover, for $m\in\set{0,\ldots,M-1}$:
\begin{align*}
    \lefteqn{P(\sgeorv_{\res_2} = n\res_1 + m\res_2 ~|~\sgeorv_{\res_1}=n\res_1)}\\
    &= P(n\res_1 + m\res_2\leq \exprv<n\res_1 + m\res_2+\res_2 ~|~n\res_1 \leq \exprv < n\res_1 + \res_1)\\
    &= \frac{\int_{n\res_1 +m\res_2}^{n\res_1+m\res_2+\res_2} \frac{1}{\beta} e^{-\frac{s}{\beta}}~ds}{\int_{n\res_1}^{n\res_1+\res_1} \frac{1}{\beta} e^{-\frac{s}{\beta}}~ds}
    = \frac{e^{-\frac{(n\res_1 +m\res_2)}{\beta}}(1-e^{-\frac{\res_2}{\beta}})}{e^{-\frac{ n\res_1}{\beta}}(1-e^{-\frac{\res_1}{\beta}})}\\
    &=\frac{(1-e^{-\frac{\res_2}{\beta}})}{(1-e^{-\frac{\res_1}{\beta}})} \cdot e^{-\frac{m\res_2}{\beta}}\quad (\textrm{let } p=1-e^{-\frac{\res_2}{\beta}})\\
    &=\frac{p(1-p)^m}{1-(1-p)^M}
\end{align*}
Thus from Lemme~\ref{lem:trunc_geo},
this is the probability mass function of a truncated geometric distribution and is independent of the value of $n\gamma_1$.
Since $\sgeorv_{\res_1}$ is a rounded down version of $\sgeorv_{\res_2}$, this also means that the distribution of $\sgeorv_{\res_2}-\sgeorv_{\res_1}$ is this same truncated geometric. Thus we have $\sgeorv_{\res_2}=\sgeorv_{\res_1} + \res_2 (\georv \bmod M)$ where $\georv$ is $\geom(1-e^{-\frac{\res_2}{\beta}})$.
\end{proof}

Now we are ready to establish the equivalence between Algorithm~\ref{alg:gaptopk3} and Algorithm~\ref{alg:gaptopk4}.

\begin{proof}[Proof of Theorem~\ref{thm:alg4}]
We first show that the output indices $j_1, \ldots, j_k$ from Algorithm~\ref{alg:gaptopk3} follow the same distribution as those from Algorithm~\ref{alg:gaptopk4}. Note that $j_1, \ldots, j_k$ are determined by $\widehat{q}_i^{\,(t)}$ in Algorithm~\ref{alg:gaptopk3} and by $\widecheck{q}_i^{\,(t)}$ in Algorithm~\ref{alg:gaptopk4} respectively. Therefore, it suffices to show that for all $t\geq 0$, the set of random variables $(\widehat{q}_1^{\,(0)}, \ldots, \widehat{q}_n^{\,(0)},\ldots, \widehat{q}_1^{\,(t)}, \ldots, \widehat{q}_n^{\,(t)})$ in Algorithm~\ref{alg:gaptopk3} and $(\widecheck{q}_1^{\,(0)}, \ldots, \widecheck{q}_n^{\,(0)},\ldots, \widecheck{q}_1^{\,(t)}, \ldots, \widecheck{q}_n^{\,(t)})$ in Algorithm~\ref{alg:gaptopk4} have the same joint distribution. 

For this part, it helps to view the variables as existing for all $t\geq 0$ before the algorithm even runs, but the algorithm only looks at the ones it needs (i.e., if the algorithm broke all ties for $t=t^*$, it does not look at $\widehat{q}_i^{\,(t^*+1)}$ or $\widecheck{q}_i^{\,(t^*+1)}$ even though they exist.

Since for $i_1\neq i_2$, the set of random variables $\set{\widehat{q}_{i_1}^{\,(0)},\ldots, \widehat{q}_{i_1}^{\,(t)}}$ (resp. $\set{\widecheck{q}_{i_1}^{\,(0)},\ldots, \widecheck{q}_{i_1}^{\,(t)}}$) is independent of $\set{\widehat{q}_{i_2}^{\,(0)},\ldots, \widehat{q}_{i_2}^{\,(t)}}$ (resp. $\set{\widecheck{q}_{i_2}^{\,(0)},\ldots, \widecheck{q}_{i_2}^{\,(t)}}$), we have
\begin{align*}
    P(\widehat{q}_{1}^{\,(0)}, \ldots, \widehat{q}_{n}^{\,(0)}, \ldots, \widehat{q}_{1}^{\,(t)}, \ldots, \widehat{q}_{n}^{\,(t)})&=\prod_{i=1}^{n}P(\widehat{q}_{i}^{\,(0)},\ldots,\widehat{q}_{i}^{\,(t)})\\
    P(\widecheck{q}_{1}^{\,(0)}, \ldots, \widecheck{q}_{n}^{\,(0)}, \ldots, \widecheck{q}_{1}^{\,(t)}, \ldots, \widecheck{q}_{n}^{\,(t)})&=\prod_{i=1}^{n}P(\widecheck{q}_{i}^{\,(0)},\ldots,\widecheck{q}_{i}^{\,(t)})
\end{align*}

Thus it suffices to show that $\forall i, t$, $P(\widehat{q}_{i}^{\,(0)},\ldots,\widehat{q}_{i}^{\,(t)})=P(\widecheck{q}_{i}^{\,(0)},\ldots,\widecheck{q}_{i}^{\,(t)}).$
Recall that in Algorithm~\ref{alg:gaptopk3}, $\widetilde{q}_{i} = q_i+\exprv$ where $\exprv\sim \expo(2k/\epsilon)$ is random noise from the exponential distribution. Furthermore, we have $\widehat{q}_{i}^{\,(t)}=\down[{\res_t}]{\widetilde{q}_i} = q_i + \down[{\res_t}]{\exprv}$ since by assumption $q_i$ is a multiple of $\tgtres=M^t\res_t$ (and hence also a multiple of $\res_t$). 
On the other hand, in Algorithm~\ref{alg:gaptopk4} we have $\widecheck{q}_{i}^{\,(t)} =q_i+ \res_0\georv_0+\res_1(\georv_1 \bmod M )\ldots+\res_t(\georv_t \bmod M)$. Let $\sgeorv_{\res_t}=\down[{\res_t}]{\exprv}$ and $\sgeorv'_{\res_t}=\res_0\georv_0 +\sum_{j=1}^t\res_j(\georv_j \bmod M)$. Then we just need to show that $P(\sgeorv_{\res_0},\ldots,\sgeorv_{\res_t})=P(\sgeorv'_{\res_0},\ldots,\sgeorv'_{\res_t})$. We do this by induction on $t$.

The base case for $t=0$ is simple. By Lemma~\ref{lem:exp_rounddown}, $\sgeorv_{\res_0}=\down[\res_0]{\exprv}$ follows the $\res_0$-scaled geometric distribution with success probability $p_0=1-e^{-\frac{\epsilon\res_0}{2k}}$, which is the same as $\sgeorv'_0=\res_0\georv_0$.

Next, we assume (as an inductive hypothesis) that $P(\sgeorv_{\res_0},\ldots,\sgeorv_{\res_t}) = P(\sgeorv'_{\res_0},\ldots,\sgeorv'_{\res_t})$ and proceed to show that $P(\sgeorv_{\res_0},\ldots,\sgeorv_{\res_{t+1}}) = P(\sgeorv'_{\res_0},\ldots,\sgeorv'_{\res_{t+1}})$.

Note that by Lemma~\ref{lem:exp_rounddown_refine} the values of $\sgeorv_{\res_{0}}, \ldots, \sgeorv_{\res_{t-1}}$ are uniquely determined by $\sgeorv_{\res_{t}}$ and so $P(\sgeorv_{\res_{t+1}}\mid \sgeorv_{\res_0}, \ldots, \sgeorv_{\res_t})=P(\sgeorv_{\res_{t+1}}\mid \sgeorv_{\res_t})$.

Also, $0\leq \sgeorv'_{\res_t} - \sgeorv'_{\res_{t-1}} = \res_t(\georv_t \bmod M) < M\res_t=\res_{t-1}$, so $\sgeorv'_{\res_{t-1}}=\down[\res_{t-1}]{\sgeorv'_{\res_{t-1}}}=\down[\res_{t-1}]{\sgeorv'_{\res_t}}$. This means that 
the values of $\sgeorv'_{\res_{0}}, \ldots, \sgeorv'_{\res_{t-1}}$ are uniquely determined by $\sgeorv'_{\res_{t}}$ and so
$P(\sgeorv'_{\res_{t+1}}\mid \sgeorv'_0, \ldots, \sgeorv'_{\res_t})=P(\sgeorv'_{\res_{t+1}}\mid \sgeorv'_{\res_t})$.


Therefore, it suffices to show that $P(\sgeorv_{\res_{t+1}}\mid \sgeorv_{\res_t})= P(\sgeorv'_{\res_{t+1}}\mid \sgeorv'_{\res_t})$. 
Since $\res_t=M\res_{t+1}$, by Lemma~\ref{lem:exp_rounddown_refine} we have that $\sgeorv_{\res_{t+1}} = \sgeorv_{\res_t} + \res_{t+1}(\georv\bmod M)$ where $\georv\sim \geom(1-e^{-\frac{\epsilon\res_{t+1}}{2k}})$. By definition we have $\sgeorv'_{\res_{t+1}}=\sgeorv'_{\res_t}+\res_{t+1}(\georv_{t+1} \bmod M)$ with $\georv_{t+1}\sim\geom(1-e^{-\frac{\epsilon\res_{t+1}}{2k}})$. Thus they have the same distribution.

Lastly we show that the output gaps $g_1,\ldots,g_k$ from the two algorithms follow the same distribution. In Algorithm~\ref{alg:gaptopk3}, the gaps are computed using $g_i=\down[\tgtres]{\widetilde{q}_{j_i} - \widetilde{q}_{j_{i+1}}}=q_{j_i}-q_{j_{i+1}}+ \down[\tgtres]{X_{j_i}-X_{j_{i+1}}} $ (since the $q_i$ are multiples of $\tgtres$). To describe the gap computation in Algorithm~\ref{alg:gaptopk4}, we need the following notation:
\begin{itemize}
\item $\pi$ is a uniformly random permutation on $1,\dots, n$ (total number of queries).
\item $\widecheck{\delta}^\pi_{ij}=-1$ if $\pi(i)<\pi(j)$ and 0 otherwise.
\item $\sgeorv'_{\res_t,i}$ is the noise in the $i^\text{th}$  noisy query at resolution $\gamma_t$. I.e., $\sgeorv'_{\res_t,i}=\widecheck{q}_{i}^{(t)} - q_{i}$ for the value of $t$ (tie breaking iteration number) at the gap calculation step.
\end{itemize}
Then the gaps in Algorithm \ref{alg:gaptopk4} are computed as $\down[\tgtres]{\widecheck{q}_{j_i}^{\,(t)}-\widecheck{q}_{j_{i+1}}^{\,(t)}-\res_t\widecheck{\delta}^\pi_{j_i,j_{i+1}}}=q_{j_i}-q_{j_{i+1}}+\down[\tgtres]{\sgeorv'_{\res_t,j_i}-\sgeorv'_{\res_t,j_{i+1}}-\res_t\widecheck{\delta}^\pi_{j_i,j_{i+1}}}$, where $j_i$ is the index of the query with the $i^\text{th}$ largest noisy answer based on the resolution $\gamma_t$ (when all ties have been broken).  Note Algorithm \ref{alg:gaptopk4} used a permutation over $1,\dots, k+1$ and we are using a permutation over $1,\dots, n$ here. This is completely equivalent because $\widecheck{\delta}^\pi_{i,j}$ is determined by how the permutation re-orders the numbers $1,\dots, n$ (each ordering is equally likely);  restricting the ordering to a subset of $1,\dots, n$ (i.e., $j_1,\dots, j_{k+1}$) still results in a uniformly random ordering on the subset.

Next we note that by Lemma~\ref{lem:exp_rounddown_refine}, $\down[\tgtres]{X_{j_i}-X_{j_{i+1}}}=\down[\tgtres]{\down[\res_t]{X_{j_i}-X_{j_{i+1}}}}$. Also, from the proof above, we know that the $\sgeorv'_{\res_t,i}$ variables (for all $i$) follow the same joint distribution as the $\down[\res_t]{X_i}$'s.
Therefore, we just need to show that $\down[\res_t]{X_{i}-X_{j}}$ and $\down[\res_t]{X_{i}}-\down[\res_t]{X_{j}}-\res_t\widecheck{\delta}^\pi_{i,j}$ follow the same joint distribution over all $i,j$ conditioned on values for all of  the $\down[\gamma_t]{X_i}$ for all $i$ (since that is what is used to determine which queries to return and is the information available to the algorithm right before the gap calculation).
This follows from the next lemma.
\end{proof}

\begin{lem}\label{roundinglem}
For any two numbers $X_1, X_2$, the following is true:
\begin{align*}
\down[\res_t]{\exprv_i-\exprv_{j}}&=\down[\res_t]{\exprv_i}-\down[\res]{\exprv_j}-
\begin{cases}
0 &\quad \textrm{if } X_i-\down[\res_t]{X_i}\geq X_j-\down[\res_t]{X_j}\\
\res_t & \quad \textrm{otherwise }\\
\end{cases}
\end{align*}
Furthermore, if $\exprv_1, \ldots,\exprv_{n}$ are i.i.d. exponential random variables and if $\pi$ is a random permutation of $1,\dots, n$ chosen uniformly at random, then the joint distribution of all the comparisons $X_i-\down[\res_t]{X_i}\geq X_j-\down[\res_t]{X_j}$ for all $i,j$ conditioned on knowledge of all of the $\down[\res_t]{X_i}$ for all $i$ is the same as the joint distribution of the comparisons $\pi(i) \geq \pi(j)$ for all $i,j$.
\end{lem}
\begin{proof}
First, $X_i = \down[\res_t]{X_i} + (X_i-\down[\res_t]{X_i})$ so
\begin{align*}
\down[\res_t]{X_i-X_j} &= \down[\res_t]{\down[\res_t]{X_i} + (X_i-\down[\res_t]{X_i}) - \down[\res_t]{X_j} - (X_j-\down[\res_t]{X_j})}\\
&=\down[\res_t]{\Big(\down[\res_t]{X_i}-\down[\res_t]{X_j}\Big) + (X_i-\down[\res_t]{X_i}) - (X_j-\down[\res_t]{X_j})}\\
&=\down[\res_t]{X_i}-\down[\res_t]{X_j} + \down[\res_t]{(X_i-\down[\res_t]{X_i}) - (X_j-\down[\res_t]{X_j})}
\end{align*}
Now, since $\res_t>X_i-\down[\res_t]{X_i}\geq 0$ (and similarly for $j$), then if $X_i-\down[\res_t]{X_i}\geq X_j-\down[\res_t]{X_j}$ we have $\res_t > X_i-\down[\res_t]{X_i} - ( X_j-\down[\res_t]{X_j})\geq  0$ and so rounding it down to the nearest multiple of $\res_t$ makes it 0.

On the other hand, if $X_i-\down[\res_t]{X_i}< X_j-\down[\res_t]{X_j}$, then $0 < X_i-\down[\res_t]{X_i}- (X_j-\down[\res_t]{X_j})\leq -\res_t$ and so rounding it down to the nearest multiple of $\res_t$ makes it $-\res_t$. This proves the first part.

For the second part, Let $\exprv_1,\dots, \exprv_n$ be i.i.d. exponential random variables. Let $\texprv_i=\exprv_i-\down[\res_t]{\exprv_i}$. 
Then from Lemma~\ref{lem:reverse_exp_rounddown} we know that $\texprv_1, \ldots, \texprv_n \mid \down[\res_t]{\exprv_1},\ldots, \down[\res_t]{\exprv_n}$ follow \emph{independent} identical truncated exponential distribution on $[0,\res)$. 
Therefore, any ordering of the $\texprv_i$ (given all of the $\down[\res_t]{\exprv_j}$) is equally likely and so has the same distribution as an ordering given by a uniformly random permutation.
\end{proof}

\section{Experiments}\label{sec:experiments}
We next evaluate the performance of the floating-point secure implementation.
We implemented our algorithms in Python. All experiments are performed on an $\text{Intel}^{\text{\textregistered}}$ $\text{Core}^{\text{\texttrademark}}$ i9-10900X @ 3.7GHz CPU machine with 64 GB memory. In all experiments, we set the privacy budget $\epsilon$ to be 1 and the target precision $\tgtres$ is set to be 0.1, meaning all noisy gap values are rounded to the first decimal place. The precision increment factor $M$ is set to 10. Thus if a precision of 0.1 is not enough to break all ties among top noisy queries, the algorithm switches precision 0.01 during tie-breaking, then 0.001 if necessary, etc.

\subsection*{Datasets}
Since the performance of the algorithm can be affected by the distribution of query answers (which affect the probability of ties in the noisy query answers), we evaluate the implementation on two real datasets BMSPOS and Kosarak from \cite{lyu2017understanding}, and a synthetic dataset T40I10D100K \cite{agrawal1994fast}. These datasets are collections of transactions (each transaction is a set of items). 
In the experiments, each query is associated with an item and the value of the query is the number of transactions the associated item appears in. The statistics of the datasets are listed below. 

\begin{table}[ht]
\caption{Statistics of datasets\label{tab:datasets}}
\begin{center}
\begin{tabular}{c c c} 
\toprule
\textbf{Dataset} & \textbf{\# of Records} & \textbf{\# of Unique Items} \\
\midrule
T40I10D100K & 100,000 & 942 \\
BMS-POS & 515,597 & 1,657 \\
Kosarak & 990,002 & 41,270\\
\bottomrule
\end{tabular}
\end{center}
\end{table}

\subsection*{Performance}
The first set of experiments compare the running times of  (1) ``Secure'': the  unoptimized  secure sampling algorithm (i.e., when a tie occurs, the resolution  of all noisy query answers is increased); (2) ``Opt. Secure'': the optimized version that uses early query pruning;  (3)  the non-secure Python implementation of Algorithm~\ref{alg:gaptopk1} which uses exponential noise as implemented in the NumPy library. The results are shown in Table~\ref{tab:time}. For $k=25, 50, 100, 200, 400$ and $800$, we run each of the three algorithms for 1000 times and report the average running time on all three datasets. 

\begin{table}[ht]
\caption{Running time (milliseconds)\label{tab:time}}
\begin{center}
\begin{tabular}{ c|c|c c c c c c} 
\toprule
Dataset & Algorithm & $k$=25 & $k$=50 & $k$=100 & $k$=200 & $k$=400 &$k$=800\\
\midrule
\multirow{3}{6em}{T40I10D100K\\$n=942$} & Secure & 10.53 & 10.80 & 11.64 & 13.95 & 19.80 & 25.89 \\ 
{} & Opt. Secure & 10.59 & 10.64 & 10.83 & 11.51 & 14.74 & 23.70 \\
{} &  Baseline & 2.25 & 2.24 & 2.27 & 2.32 & 2.38 & 2.52\\
\midrule
\multirow{3}{6em}{BMS-POS\\$n=1657$} &  Secure & 18.51 & 18.51 & 19.03 & 22.11 & 35.21 & 48.72 \\ 
{} &  Opt. Secure & 18.43 & 18.47 & 18.56 & 19.11 & 22.85 & 33.09 \\
{} &  Baseline & 3.92 & 3.91 & 3.96 & 3.99 & 4.06 & 4.20\\
\midrule
\multirow{3}{6em}{Kosarak\\$n=41270$} &  Secure & 458.22 & 459.45 & 482.02 & 575.38 & 976.72 & 1223.04\\ 
{} & Opt. Secure & 455.32 & 455.17 & 454.89 & 455.59 & 459.73 & 471.30\\
{} &  Baseline & 98.71 & 97.53 & 97.38 & 98.69 & 98.95 & 99.03\\
\bottomrule
\end{tabular}
\end{center}
\end{table}

For relatively small values of $k$, both versions of the secure algorithm are approximately 4.7$\times$ slower than the insecure algorithm, but this overhead is negligible because the overall runtime is in tens of milliseconds. 
When $k$ is large, the unoptimized implementation becomes visibly slower  because large values of $k$ will make ties more likely to happen.

\subsection*{Running Time Breakdown} Next we focus on the optimized secure implementation and analyze its potential bottleneck. We separate Algorithm~\ref{alg:gaptopk4} into three parts. The first part (from Line~\ref{line:part1begin} to Line~\ref{line:part1end}) is to initially identify the possible top $k$ noisy queries. Because of the possibility of the existence of ties among the chosen queries, the next part (from Line~\ref{line:part2begin} to Line~\ref{line:part2end}) use a loop to iteratively increase noise precision until all ties are resolved. Finally, the last part (from Line~\ref{line:part3begin} to Line~\ref{line:part3end}) is to obtain numeric gaps estimates among chosen queries at target precision. We instrument the algorithm with timing instructions 
and report the average time spent on the three parts over 100 iterations. The results are summarised in Table~\ref{tab:prof}.

\begin{table}[ht]
\caption{Detailed time for the optimized secure algorithm (milliseconds)\label{tab:prof}}
\begin{center}
\begin{tabular}{ c|c|c @{\hspace{0.8\tabcolsep}} c@{\hspace{0.8\tabcolsep}} c@{\hspace{0.8\tabcolsep}} c@{\hspace{0.8\tabcolsep}} c@{\hspace{0.8\tabcolsep}} c} 
\toprule
Dataset & Time spent to & $k$=25 & $k$=50 & $k$=100 & $k$=200 & $k$=400 &$k$=800\\
\midrule
\multirow{3}{6em}{T40I10D100K\\$n=942$} & Possible top-$k$& 10.921 & 10.530 & 10.500 & 10.457 & 10.527 & 10.555\\ 
{} &  Resolve ties & 0.005 & 0.017 & 0.190 & 0.622 & 3.813 & 11.573 \\
{} &  Find gaps & 0.049 & 0.081 & 0.154 & 0.300 & 0.599 & 1.192\\
\midrule
\multirow{3}{6em}{BMS-POS\\$n=1657$} & Possible top-$k$ & 18.475 & 18.294 & 18.261 & 18.199 & 18.159 & 18.188\\ 
{} &  Resolve ties & 0.005 & 0.023 & 0.057 & 0.288 & 4.092 & 14.187 \\
{} &  Find gaps & 0.045 & 0.081 & 0.154 & 0.302 & 0.598 & 1.181\\
\midrule
\multirow{3}{6em}{Kosarak\\$n=41270$} & Possible top-$k$ & 461.56 & 457.82 & 457.371 & 458.721 & 456.256 & 459.552 \\ 
{} &  Resolve ties & 0.008 & 0.012 & 0.074 & 0.744 & 5.077 & 15.586\\
{} &  Find gaps & 0.080 & 0.114 & 0.186 &  0.340 & 0.618 & 1.218\\
\bottomrule
\end{tabular}
\end{center}
\end{table}

From the results there are several observations. 
First, the time spent to initially identify the possible top $k$ queries is determined by $n$ (total number of queries). This is consistent with our expectation because the algorithm needs to add noise to all $n$ queries and then sort the noisy queries, which takes $O(n\log n)$ time (or $O(kn)$ when searching directly for the top $k+1$ queries without sorting). 
Second, the time needed to resolve ties among top $k$ queries is roughly $O(k\log^2 k)$. This is because each tie-breaking iteration runs in $O(k\log k)$ time (adding noise to roughly $k$ queries and sorting them). Although more queries could result in more ties, a tie that is resolved in an iteration will not reappear in a later iteration. Thus the expected number of ties among the top $k$ queries decrease exponentially with the number of iterations, and the expected number of iterations needed to break all ties is roughly $\log k$. 
Third, the time to get gaps among chosen queries is proportional to $k$, and is generally negligible. We remark that these numbers are consistent with the numbers reported in Table~\ref{tab:time}, in that the time spent on the three parts of the optimized algorithm sum up to roughly the time reported in Table~\ref{tab:time} (middle row, Opt. Secure). 

\subsection*{Time Spent in Different Sampling Commands} Last, we use a Python profiler (cProfile) to track the number of samples drawn from different sampling primitives during a single run of the optimized algorithm. We run our optimized algorithm on the Kosarak dataset ($n = 41270$) with $k=800$. Recall that the optimized algorithm uses noise drawn from the geometric distribution (see Algorithm~\ref{alg:geometric} in the Appendix), which subsequently draws samples from the Bernoulli and uniform distributions over integers (see Algorithm 1 in \cite{Canonne_Kamath_Steinke_2022}). The number of evocations for each sampling command and the total time spent on it (including all subfunction calls) are reported in Table~\ref{tab:cprof}. We also include the sorting function in this table as it clearly indicates how many times the tie breaking loop is run. Recall that the optimized algorithm first add geometric noise to all $n = 41270$ queries and call the sorting function to identify possible top queries. Then for each tie breaking loop iteration, additional geometric noise is drawn for each query in the pool of relevant queries to increase query answer resolution, and the sorting is called exactly once.

\begin{table}[ht]
\caption{Statistics on function invocations during a single execution\label{tab:cprof}}
\begin{center}
\begin{tabular}{l c r} 
\toprule
\textbf{Subroutine name} & \textbf{Total time spent} & \textbf{Number of calls} \\
\midrule
\gaptopk (optimized) & 1.208 & 1 \\
Geometric distribution sampling & 1.011 & 42,874 \\
Bernoulli sampling & 0.735 & 303,216\\
Uniform sampling over bounded integers & 0.779 & 369,613\\
Identifying top noisy queries (argsort) & 0.002 & 3\\
\bottomrule
\end{tabular}
\end{center}
\end{table}

From this table, we can see that the optimized algorithm ran 2 loop iterations to resolve ties (argsort is called 3 times). Note also that the total number of geometric noise sampled is 42874. Since there are $n=41270$ queries in the dataset, $42874-41270=1604$ additional geometric samples are used for tie breaking, $802$ in each of the two loop iteration. This means early query elimination successfully removed queries that are definitely not among the top $k=800$, leaving only $k+2=802$ (the minimum required by the algorithm) queries for further investigation. On average, each secure geometric sampling command calls 7.07 Bernoulli and 8.62 uniform sampling commands. Lastly, it can be seen from the table that most of the time is spent on securely sampling random noise from various distributions.

\section{Conclusion}\label{sec:conclusion}
In this paper, we proposed an implementation of the differentially private \gaptopk algorithm that avoids the use of floating point and hence is secure against floating point side channels. The algorithm is probabilistically equivalent to the ideal algorithm that uses continuous noise and rounds its inputs and outputs. The algorithm makes heavy use of a variety of memoryless properties of geometric and exponential random variables to dynamically determine the appropriate level of discretization for the noise and uses carefully controlled randomized rounding routines to provide the gap information.

\section*{Acknowledgment}
This work was support by NSF awards CNS 1702760 and CNS 1931686.


\bibliography{refs}
\bibliographystyle{abbrv}
\clearpage
\appendix
\section{Geometric Sampling Primitive}
$~$

\begin{algorithm}[h!]
\SetKwProg{Fn}{function}{\string:}{}
\SetKwFunction{Test}{Geometric}
\SetKwInOut{Input}{input}
\SetKwInOut{Output}{output}
\DontPrintSemicolon
\Input{Integers $s,t\geq 1$}
\Output{One sample from $\geom(1-e^{-s/t})$}
\Fn{\Test{$1-e^{-s/t}$}}{
$D \gets 0$\;
\While{$D=0$}{
$U\gets\unif\set{0,\ldots,t-1}$\;
$D\gets \bern(e^{-U/t})$~\tcp*[r]{Use Algorithm~1 in \cite{Canonne_Kamath_Steinke_2022}}
}
$V\gets 0$~\tcp*[r]{Generate $V$ from $\geom(1-e^{-1})$}
\While{$\mathtt{true}$}{
$A\gets\bern(e^{-1})$~\tcp*[r]{Use Algorithm~1 in \cite{Canonne_Kamath_Steinke_2022}}
\uIf{$A=0$}
{\textbf{Break}}
\Else{$V\gets V+1$}
}
$X\gets U+t\cdot V$~\tcp*[r]{$X$ is $\geom(1-e^{-1/t})$}
$Y\gets \floor{X/s}$~\tcp*[r]{$Y$ is $\geom(1-e^{-s/t})$}
\Return $Y$
}
\caption{Sampling from a Geometric Distribution (extracted from Algorithm 2 in  \cite{Canonne_Kamath_Steinke_2022})}
\label{alg:geometric}
\end{algorithm}

\end{document}